\documentclass[aps,prc,showpacs,floatfix]{revtex4}
\usepackage[dvips]{graphicx}
\begin{document}

\draft
\title{Proton inelastic diffraction by a black nucleus and the size of 
excited nuclei}

\author{Kei Iida,$^{1,2}$ Shinya Koide,$^1$ Akihisa Kohama,$^2$ and 
Kazuhiro Oyamatsu$^{2,3}$}
\affiliation{
$^1$Department of Natural Science, Kochi University, Kochi 780-8520, 
Japan\\
$^2$RIKEN Nishina Center, RIKEN,
2-1 Hirosawa, Wako-shi, Saitama 351-0198, Japan\\
$^3$Faculty of Human Informatics, Aichi Shukutoku University, Nagakute, 
Nagakute-cho, Aichi-gun, Aichi 480-1197, Japan}

\date{\today}

\begin{abstract}

     We systematically derive a length scale characterizing the size of a 
low-lying, $\beta$ stable nucleus from empirical data for the diffraction peak 
angle in the proton inelastic differential cross section of incident energy
of $\sim1$ GeV.  In doing so, we assume that the target nucleus in the ground 
state is a completely absorptive ``black'' sphere of radius $a$.  The cross 
section $\pi a^2$, where $a$ is determined in such a way as to reproduce the 
empirical proton diffraction peak angle in the elastic channel, is known to 
agree with empirical total reaction cross sections for incident protons to 
within error bars.  By comparing the inelastic diffraction patterns obtained in
the Fraunhofer approximation with the experimental ones, one can likewise 
derive the black sphere radius $a_l$ for the excited state with spin $l$. We 
find that for $^{12}$C, $^{58,60,62,64}$Ni, and $^{208}$Pb, the value of $a_l$ 
obtained from the inelastic channel is generally larger than the value of 
$a$ from the elastic channel and tends to increase with the excitation 
energy.  This increase is remarkable for the Hoyle state.  Finally, we 
discuss the relation between $a_l$ and the size of excited nuclei.

\end{abstract}

\maketitle

\section{Introduction}

     The size of atomic nuclei is one of the most fundamental quantities that 
characterize matter in the nuclei. It is well known for $\beta$ stable nuclei 
in the ground state thanks to systematic measurements of electron and proton 
elastic differential cross sections \cite{Bat:ANP}.  This helps clarify the 
equation of state of nuclear matter near the saturation point \cite{oyaii}.  
For excited states of $\beta$ stable nuclei, however, it is not 
straightforward to deduce the 
nuclear size, because elastic scattering off an excited target is hard to 
measure.  Alternatively, one can use proton inelastic differential cross 
sections in deducing the size of excited nuclei, but all one may know is the 
transition density, which only implicitly reflects the density distribution 
of the excited nuclei.

     Although one can in principle determine the charge and matter density 
distributions of a target nucleus from measured electron and proton elastic 
differential cross sections, it is practically time-consuming even if one 
uses optical potential models and Glauber's multiple scattering theory
in an approximate manner.  Instead of sticking to such microscopic 
derivations of the nuclear size, in Ref.\ \cite{KIO1} we constructed a 
phenomenological method for deducing the nuclear size by focusing on the peak 
angle in the proton-nucleus elastic differential cross section measured 
at proton incident energy $T_p\sim$800-1000 MeV, where the corresponding 
optical potential is strongly absorptive.  In this method, we regard a 
nucleus as a ``black'' (i.e., purely absorptive) sphere of radius $a$, 
and we determine $a$ in such a way that the first peak angle of the Fraunhofer
diffraction by a circular black disk of radius $a$ agrees with that of
the measured diffraction.  This method is reasonable as long as the
scattering is close to the limit of the geometrical optics.  This condition 
is fairly well satisfied at least for $T_p\gtrsim800$ MeV, where
the wave length of the incident proton is sufficiently shorter than $a$
even for $^4$He.  The black sphere picture is originally expected to
give a decent description of total reaction cross sections for any kind of 
incident particle that tends to be attenuated in the nuclear interiors.  
In fact we showed that for proton beams incident on stable nuclei, the
cross section of the black sphere of radius $a$ thus determined is 
consistent with the measured total reaction cross section \cite{KIO2}.  If we 
multiply $a$ by $\sqrt{3/5}$, furthermore, a ratio between the 
root-mean-square and squared off radii for a rectangular distribution, the 
result for stable nuclei of $A\gtrsim50$ shows an excellent agreement with 
the root-mean-square radius, $r_m$, of the matter density distribution as 
determined from conventional scattering theories so as to reproduce the 
overall diffraction pattern and analyzing power in the proton elastic 
scattering \cite{KIO1}.

     In this paper, we apply the black sphere picture to analyses of proton 
inelastic scattering data for $T_p\sim$1000 MeV.  Basically, this application 
is a straightforward extension of the case of proton elastic scattering, which 
is closely related to the method developed by Blair \cite{Blair} for 
alpha scattering by assuming elastic diffraction by a circular black disk of 
radius $R_0$ and inelastic diffraction by a black nucleus with small multipolar
deformations from a sphere of radius $R_0$.  The present extension is, however,
accompanied by a nontrivial choice of the inelastic diffraction peak whose 
angle is to be fitted to the empirical value and by a variation of the black 
sphere radius from the value determined from the elastic diffraction peak 
angle.  The resulting black sphere radius does not correspond to the size of 
the nucleus excited by the incident proton, but rather is related to the 
transition density and thus expected to lie between the sizes in the ground 
state and in the excited state.  Even so, as we shall see, systematic 
derivation of the black sphere radius from the inelastic channels is 
useful for predicting how the size in the excited state depends on the 
excitation energy $E_{\rm ex}$.

     In Sec.\ II we extend the black sphere approach developed for proton
elastic diffraction to the case of proton inelastic diffraction.  The results 
for the black sphere radii are illustrated in Sec.\ III.

\section{Black sphere approach}

     We begin by summarizing the black sphere approach to proton elastic
diffraction \cite{KIO1}.  The center-of-mass (c.m.) scattering angle for proton
elastic scattering is generally given by $\theta_{\rm c.m.} = 2\sin^{-1}(q/2p)$
with the momentum transfer ${\bf q}$ and the proton incident momentum in the
c.m.\ frame ${\bf p}$.  For the proton diffraction by a circular black
disk of radius $a$, we can calculate the value of $\theta_{\rm c.m.}$
at the first peak as a function of $a$.  (Here we define the zeroth
peak as that whose angle corresponds to $\theta_{\rm c.m.}=0$.)
We determine $a$ in such a way that this value of $\theta_{\rm c.m.}$
agrees with the first peak angle for the measured diffraction in
proton-nucleus elastic scattering, $\theta_M$.  The radius, $a$, and the angle,
$\theta_M$, are then related by 
\begin{equation}
   2 p a \sin(\theta_M/2) = 5.1356 \cdots.
    \label{a}
\end{equation}
By setting
\begin{equation}
   r_{\rm BS}\equiv \sqrt{3/5}a,
   \label{rbb}
\end{equation}
we found \cite{KIO1} that at $T_p\gtrsim800$ MeV, $r_{\rm BS}$, estimated for 
heavy stable nuclei of $A\gtrsim50$, is within error bars consistent with 
the root-mean-square nuclear matter radius, $r_m$, deduced from 
elaborate analyses based on conventional scattering theory.  
Thus, expression (\ref{rbb}) works as a ``radius formula.'' 
The factor $\sqrt{3/5}$ comes from the assumption that the 
nucleon distribution is rectangular; the root-mean-square radius of a 
rectangular distribution is a cutoff radius multiplied by $\sqrt{3/5}$.
For stable nuclei with $A\lesssim50$, however, the values of $r_{\rm BS}$ are 
systematically smaller than those of $r_m$ \cite{KIO2}.  The scale $a$ is 
nevertheless meaningful because the values of $\pi a^2$ for C, Sn, and Pb 
agree well with the proton-nucleus reaction cross section data for 
$T_p\gtrsim800$ MeV \cite{KIO2}.  This indicates that $a$ can be
regarded as a ``reaction radius,'' inside which the reaction with incident
protons occurs.  In a real nucleus, this radius corresponds to the radius at 
which the mean free path of incident protons is of the order of the length of 
the penetration.  We remark that even for deformed nuclei, this interpretation
works well unless the degree of deformations is extremely large.

   We now proceed to generalize the black sphere picture to the case of proton 
inelastic scattering by following a line of argument of Blair \cite{Blair}. We
assume that the final low-lying spin-$l$ excitation of a target even-even 
nucleus is characterized by small multipolar deformations of a black sphere of 
radius $a_l$.  As we will see, this radius generally differs from the value of 
$a$ obtained from the measured peak angle of elastic diffraction off the same 
target, as well as among different levels with the same $l$.  Therefore, the 
radius $a_l$ is conceptually different from the radius $R_0$ used in Ref.\ 
\cite{Blair}.  In the adiabatic approximation, which is valid for the proton 
incident energies, $T_p\gtrsim800$ MeV, of interest here, we may set the 
reaction radius as
\begin{equation}
  R_l(\theta,\phi)=a_l\left[1+\sum_{m}\alpha_{lm}Y_{lm}(\theta,\phi)\right],
   \label{R}
\end{equation}
where $\theta$ is measured with respect to the incident beam axis, and 
$\alpha_{lm}$ are the deformation parameters.  For given $\alpha_{lm}$
and ${\bf q}$, we write the scattering amplitude in the c.m.\ frame of the
proton of initial momentum ${\bf p}$ and the black nucleus as
\begin{eqnarray}
  f({\bf q};\alpha_{lm})&=&\frac{ip}{2\pi}
           \int_0^{2\pi}d\phi \int_0^\infty db b \nonumber \\ 
        & & \times e^{-i{\bf q}\cdot{\bf b}}\theta(R_l(\pi/2,\phi)-b),
  \label{f}
\end{eqnarray}
where ${\bf b}$ is the impact parameter perpendicular to ${\bf p}$.  Here we 
assume that the final proton momentum ${\bf p}+{\bf q}$ has the magnitude 
equal to that of the initial momentum $p$ because $T_p$ is far larger than 
the excitation energy $E_{\rm ex}$.  Then, the c.m.\ scattering angle
again becomes $\theta_{\rm c.m.} = 2\sin^{-1}(q/2p)$.  With the convention
that $\phi$ is measured with respect to the projected final proton
momentum onto a plane perpendicular to ${\bf p}$, the scattering amplitude
up to linear order in $\alpha_{lm}$ reads
\begin{eqnarray}
  f({\bf q};\alpha_{lm})&=&\frac{ip}{2\pi}
           \int_0^{2\pi}d\phi \left[\int_0^{a_l}   
              e^{-iqb \cos\phi}  b db \right.  \nonumber \\
            & & \left.+e^{-iqa_l \cos\phi} 
                 a_l^2 \sum_m \alpha_{lm}Y_{lm}(\pi/2,\phi)\right].
  \label{f2}
\end{eqnarray}
The first integral would give the amplitude for elastic scattering off the
excited nucleus with spin $l$:
\begin{equation}
\frac{ipa_l^2J_1(2pa_l\sin(\theta_{\rm c.m.}/2))}
{2pa_l\sin(\theta_{\rm c.m.}/2)}.
  \label{f3}
\end{equation}
Note that this amplitude is independent of $\alpha_{lm}$.  
Anyway, the first integral is beyond 
the scope of real experiments.  The second integral in Eq.\ (\ref{f2}) leads 
to the amplitude for inelastic scattering that excites a target nucleus in the
ground state to the final state of spin $l$ \cite{Blair}:
\begin{eqnarray}
& &ipa_l^2\sum_{m=-l,-l+2,\cdots}^{l} 
\left(\frac{2l+1}{4\pi}\right)^{1/2} i^l 
\frac{[(l-m)!(l+m)!]^{1/2}}{(l-m)!!(l+m)!!}
\nonumber \\ &\times& 
\alpha_{lm}J_{|m|}(2pa_l\sin(\theta_{\rm c.m.}/2)),
    \label{f4}
\end{eqnarray}
where $(2n)!!\equiv 2\cdot4\cdot6\cdot\cdot\cdot2n$.  The corresponding 
differential cross section is
\begin{eqnarray}
\frac{d\sigma}{d\Omega}(0\to l)&\propto&
\sum_{m=-l,-l+2,\cdots}^{l}
\frac{(l-m)!(l+m)!}{[(l-m)!!(l+m)!!]^2}
\nonumber \\ &&\times 
\alpha_{lm}^2 J_{|m|}^2(2pa_l\sin(\theta_{\rm c.m.}/2)),
\label{dsinela}
\end{eqnarray}
which gives the inelastic Fraunhofer diffraction pattern.

     We next determine $a_l$ by comparing the peak angle of the black sphere
inelastic diffraction as described by Eq.\ (\ref{dsinela}) with the 
corresponding empirical value $\theta_{Ml}$.  Generally, the inelastic 
diffraction pattern at forward angles is more complicated than the 
elastic one.  We thus avoid the peak that is the nearest to 
$\theta_{\rm c.m.}=0$ and consider which peak to be chosen among the rest.  
The guiding principle for this choice is simply to search 
for the peak that systematically corresponds to the first peak in the
elastic diffraction pattern just like the case of electron diffraction
\cite{Friedrich}.  We thus obtain
\begin{equation}
   2 p a_0 \sin(\theta_{M0}/2) = 7.015 \cdots,
    \label{a0}
\end{equation}
\begin{equation}
   2 p a_2 \sin(\theta_{M2}/2) = 6.783 \cdots,
    \label{a2}
\end{equation}
\begin{equation}
   2 p a_3 \sin(\theta_{M3}/2) = 8.209 \cdots,
    \label{a3}
\end{equation}
\begin{equation}
   2 p a_4 \sin(\theta_{M4}/2) = 9.617 \cdots.
    \label{a4}
\end{equation}
The values of the right side in Eqs.\ (\ref{a0})--(\ref{a4}) correspond
to one of the values of $2pa_l\sin(\theta_{\rm c.m.}/2)\equiv x$ where 
Eq.\ (\ref{dsinela}) is locally maximal.  In fact, the values of $x$ are as 
follows: For $l=0$, $x=0$, 3.831$\cdots$, 7.015$\cdots$, $\ldots$; for $l=2$,
$x=0$, 3.251$\cdots$, 6.783$\cdots$, $\ldots$; for $l=3$, $x=1.982\cdots$, 
4.605$\cdots$, 8.209$\cdots$, $\ldots$; for $l=4$, $x=0$, 3.675$\cdots$, 
5.852$\cdots$, 9.617$\cdots$, $\ldots$.  The choice of other diffraction
peaks in extracting $\theta_{Ml}$ would in many cases produce $a_l$ that 
is significantly different from $a$.  We remark that the avoidance of
the local diffraction maxima at forward angles justifies the neglect
of Coulomb effects in the present black sphere approach \cite{Blair}.

     The extraction of $\theta_{Ml}$ from the measured inelastic
differential cross sections is not always straightforward for several 
reasons.  Firstly, in the case in which the empirical data lack such 
diffraction maxima as are supposed to occur in the Fraunhofer diffraction, 
counting the empirical diffraction maxima does not work.  It is thus 
indispensable to compare the overall measured diffraction pattern 
with the elastic one whose first peak is easily distinguishable, 
before identifying the local maxima that give $\theta_{Ml}$.  Secondly, 
even after the local maxima are successfully identified, we need to
allow for various uncertainties accompanying determination of $\theta_{Ml}$.

     As in the case of elastic scattering \cite{KIO1}, we basically determine 
the values of $\theta_{Ml}$ from the scattering angle that gives the maximum 
value of the cross section among discrete data near the identified diffraction 
maximum.  Some examples are shown in Fig.\ 1.
When the data stagger in such a way that the scattering angle that 
gives the local maximum is significantly away from the diffraction peak 
position deduced from the overall plot of the data, we select the data point 
that obviously seems the closest to the peak position. Such determination of 
$\theta_{Ml}$ is accompanied by uncertainties in the measured scattering angle 
and systematic errors that are dependent on the way of deducing the peak 
position.  The uncertainties in the measured angle, which are due mainly to 
the absolute angle calibration, are typically of order or smaller than 
$\pm0.03$ deg \cite{Ray:PRC18} for existing data for proton scattering off 
stable nuclei.  On the other hand, the systematic errors can be estimated by 
assuming that the true peak is located in the region enclosed by the two
data points that are the closest neighbors of the selected data point. 
The systematic errors thus estimated dominate the error bars of $\theta_{Ml}$, 
and hence we will ignore the uncertainties in the measured angle.  

\begin{figure}[t]
\begin{center}
\includegraphics[width=17cm]{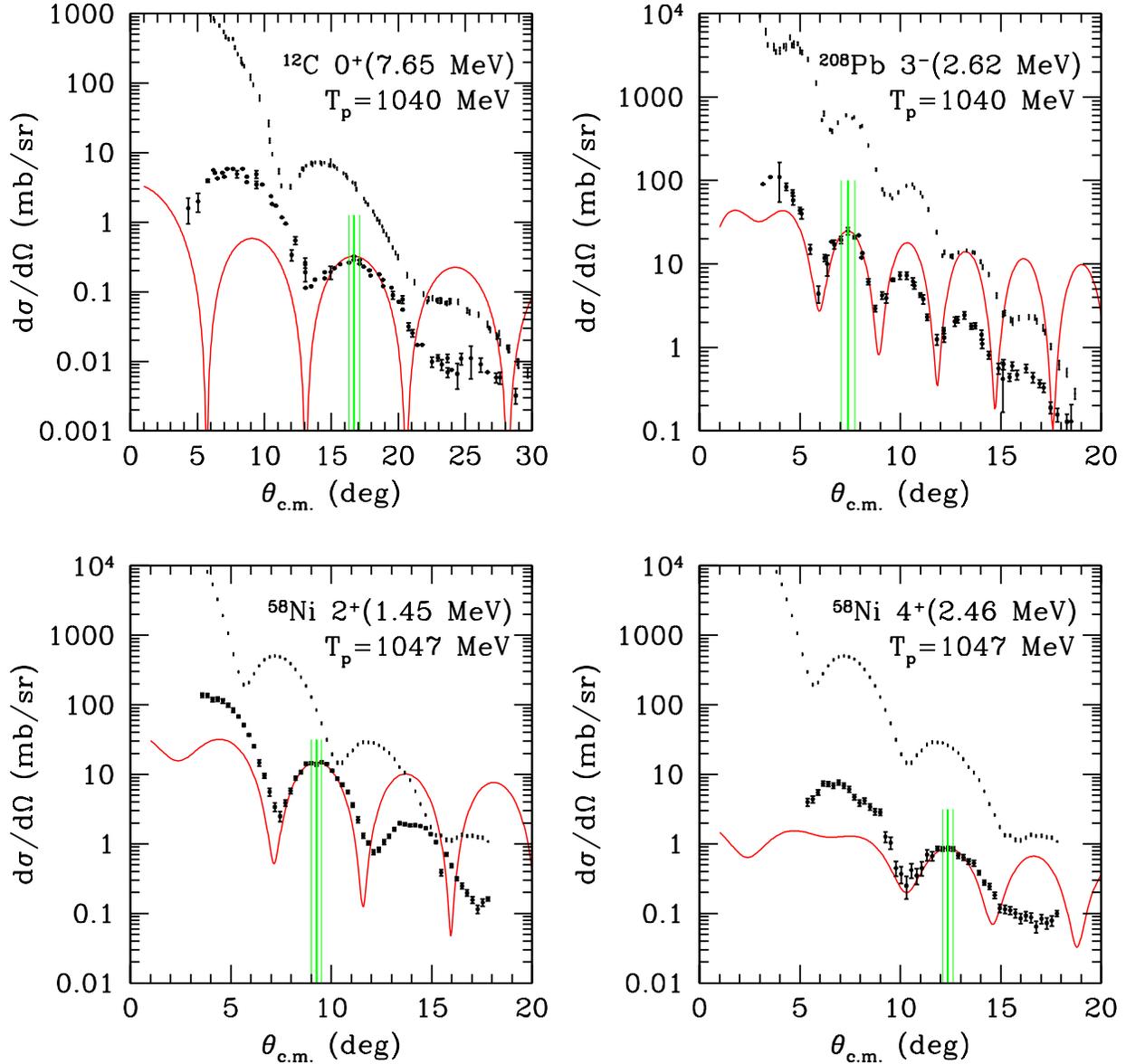}
\end{center}
\caption{
(Color online) The inelastic differential cross section calculated from 
the Fraunhofer diffraction formula (\ref{dsinela}) for $p$-$^{12}$C and 
$p$-$^{208}$Pb ($T_p=1040$ MeV) as well as $p$-$^{58}$Ni ($T_p=1047$ MeV).  The
experimental data (large dots) are taken from Refs.\ \cite{Bertini,Lombard};
for comparison, the elastic data (small dots) are also plotted.  In each 
panel, the vertical lines denote the peak angle $\theta_{Ml}$ (including error 
bars) at which we fit the calculated peak angle to the empirical one, and the 
normalization of Eq.\ (\ref{dsinela}) is set in such a way that the value of 
the differential cross section at $\theta_{Ml}$ agrees with the empirical one.
}
\end{figure}

\section{Black sphere radii}

      We finally obtain $a_l$ from the determined $\theta_{Ml}$ via Eqs.\ 
(\ref{a0})--(\ref{a4}).  For comparison, we likewise determine $\theta_M$ and 
then $a$ via Eq.\ (\ref{a}).  We remark that a part of the errors of $a_l$ 
and $a$ that come from uncertainties in the proton incident energy, which are 
typically a few MeV \cite{Ray:PRC18}, are negligible.  The results for $a$ and 
$a_l$ obtained from empirical scattering data off $^{12}$C, $^{58,60,62,64}$Ni,
and $^{208}$Pb at proton incident energy of about 1 GeV \cite{Bertini,Lombard} 
are listed in Table I.  In collecting the data, we have made access to the 
Experimental Nuclear Reaction Data File (EXFOR) \cite{IAEA}.  In the absence of
the inelastic scattering data for the $4^+$ state of $^{12}$C and the $2^+$ and
$4^+$ states of $^{208}$Pb, the corresponding black sphere radii are 
unavailable.  Note that $a_3$ is unavailable for $^{12}$C despite the presence 
of the data for the $3^-$ state of $^{12}$C, because the corresponding peak is 
missing in the measured differential cross section.

\begin{table}
\caption{Black sphere radii obtained from proton elastic and inelastic 
scattering data.
}
\begin{center}
\begin{tabular}{ccccc}  % {|c|c|c|c|c|}
\hline
Target & Final state & $E_{\rm ex}$ (MeV) & $T_p$ (MeV) & $a$ or $a_{l}$ (fm)\\
\hline
$^{12}$C & g.s.  & 0 & 1040 & 2.75$\pm$0.06  \\
         & $2^+$ & 4.44 & 1040 & 2.70$\pm$0.06  \\
         & $0^+$  & 7.65 & 1040 & 3.20$\pm$0.07  \\
\hline
$^{58}$Ni & g.s.  & 0 & 1047 & 4.79$\pm$0.18  \\
         & $2^+$ & 1.45 &1047 & 4.91$\pm$0.14  \\
         & $4^+$  & 2.46 & 1047 & 5.22$\pm$0.11  \\
         & $3^-$  & 4.47 & 1047 & 5.09$\pm$0.13  \\
\hline
$^{60}$Ni & g.s.  & 0 & 1047 & 4.78$\pm$0.18  \\
         & $2^+$ & 1.33 &1047 & 4.91$\pm$0.14  \\
         & $4^+$  & 2.50 & 1047 & 5.22$\pm$0.12  \\
         & $3^-$  & 4.04 & 1047 & 5.09$\pm$0.13  \\
\hline
$^{62}$Ni & g.s.  & 0 & 1047 & 4.96$\pm$0.19  \\
         & $2^+$ & 1.17 &1047 & 5.05$\pm$0.15  \\
         & $4^+$  & 2.34 & 1047 & 5.22$\pm$0.12  \\
         & $3^-$  & 3.76 & 1047 & 5.22$\pm$0.13  \\
\hline
$^{64}$Ni & g.s.  & 0 & 1047 & 4.96$\pm$0.19  \\
         & $2^+$ & 1.35 &1047 & 5.04$\pm$0.15  \\
         & $4^+$  & 2.61 & 1047 & 5.22$\pm$0.11  \\
         & $3^-$  & 3.55 & 1047 & 5.09$\pm$0.13  \\
\hline
$^{208}$Pb & g.s.  & 0 & 1040 & 7.49$\pm$0.35  \\
         & $3^-$  & 2.62 & 1040 & 7.29$\pm$0.35  \\
\hline
\end{tabular}
\end{center}
\end{table}

\begin{figure}
\begin{center}
 \begin{minipage}{.45\linewidth}
\includegraphics[width=8.5cm]{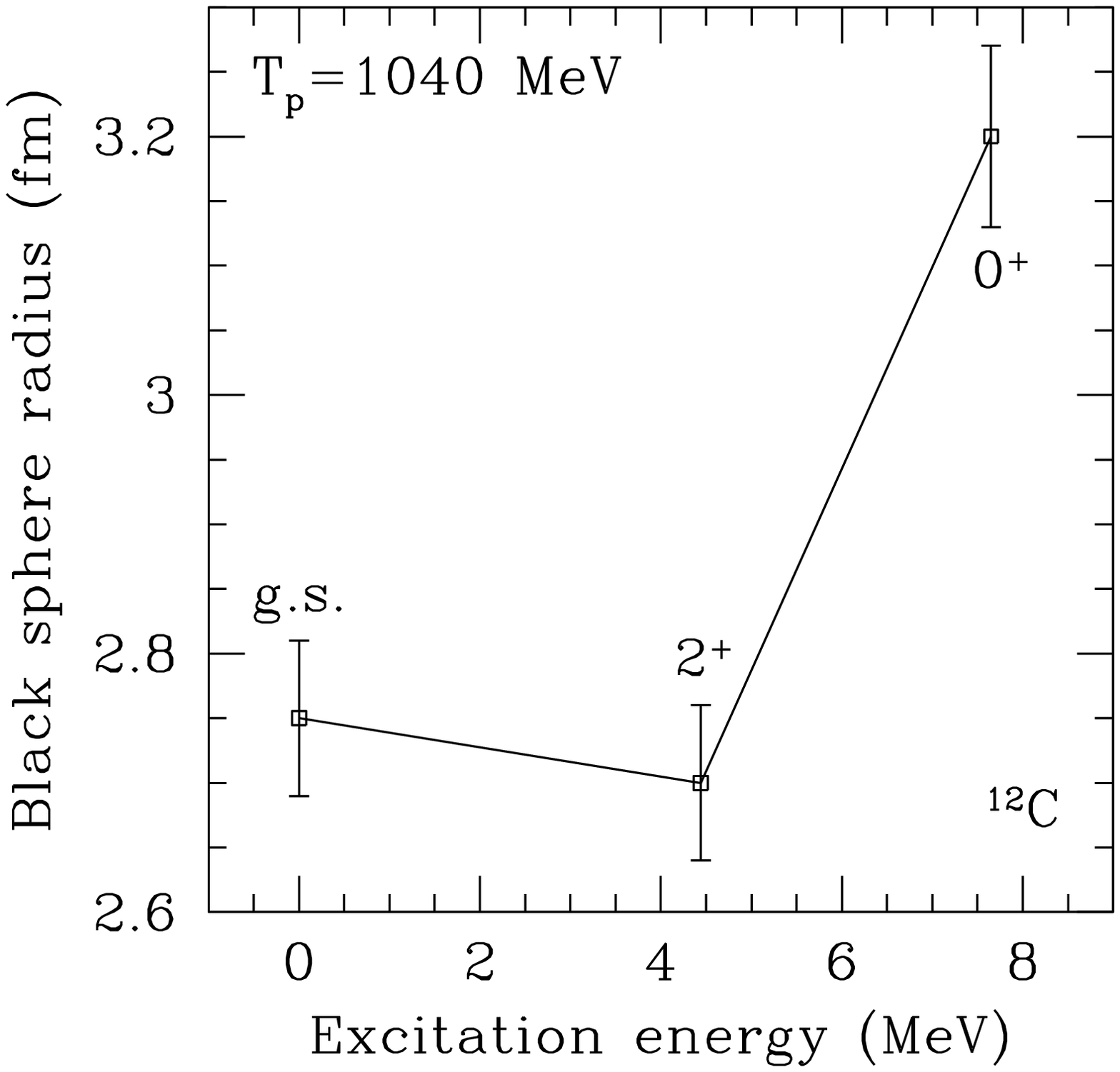}
 \end{minipage}
 \begin{minipage}{.45\linewidth}
\includegraphics[width=8.5cm]{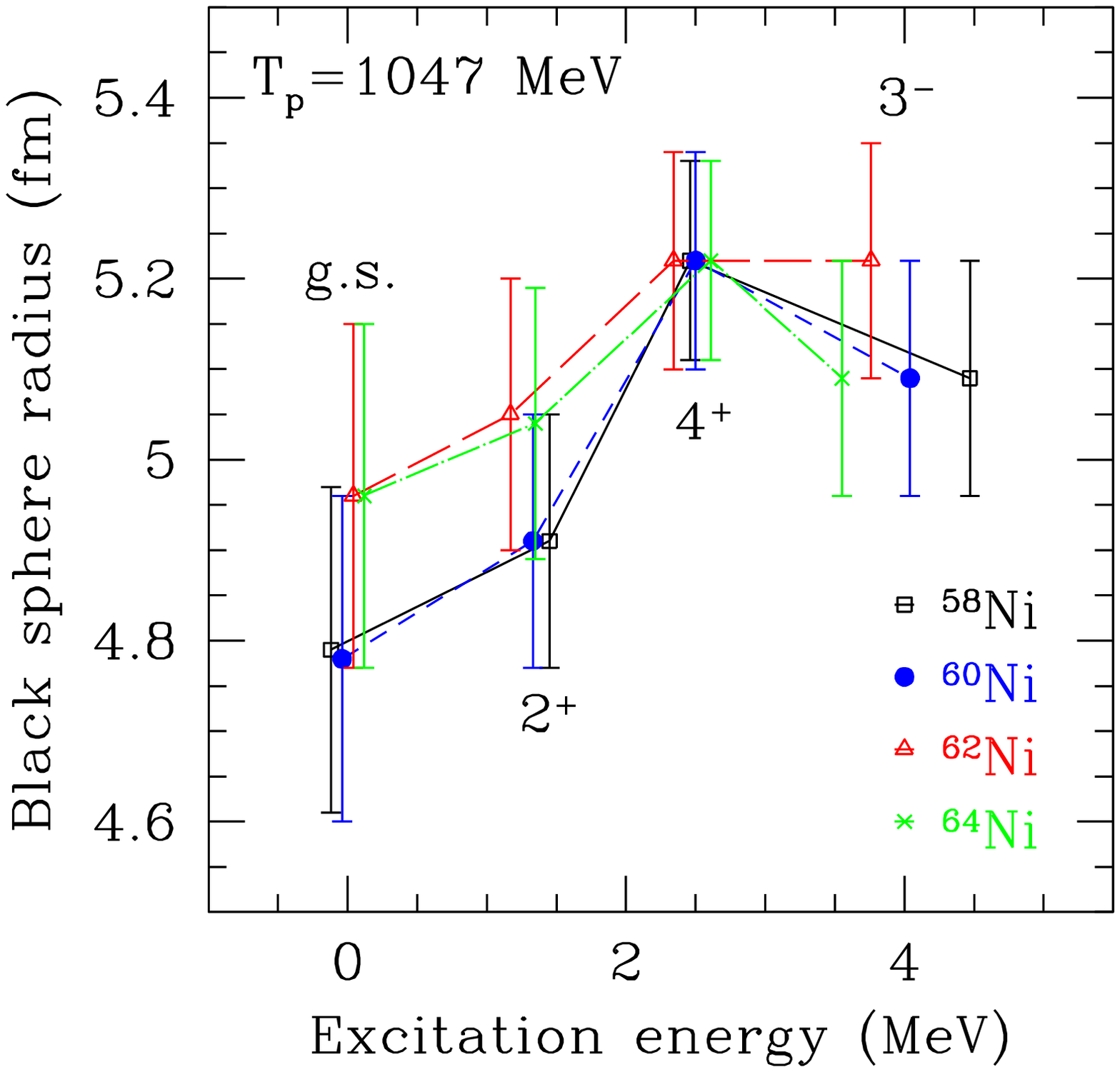}
 \end{minipage}
 \begin{minipage}{.45\linewidth}
\includegraphics[width=8.5cm]{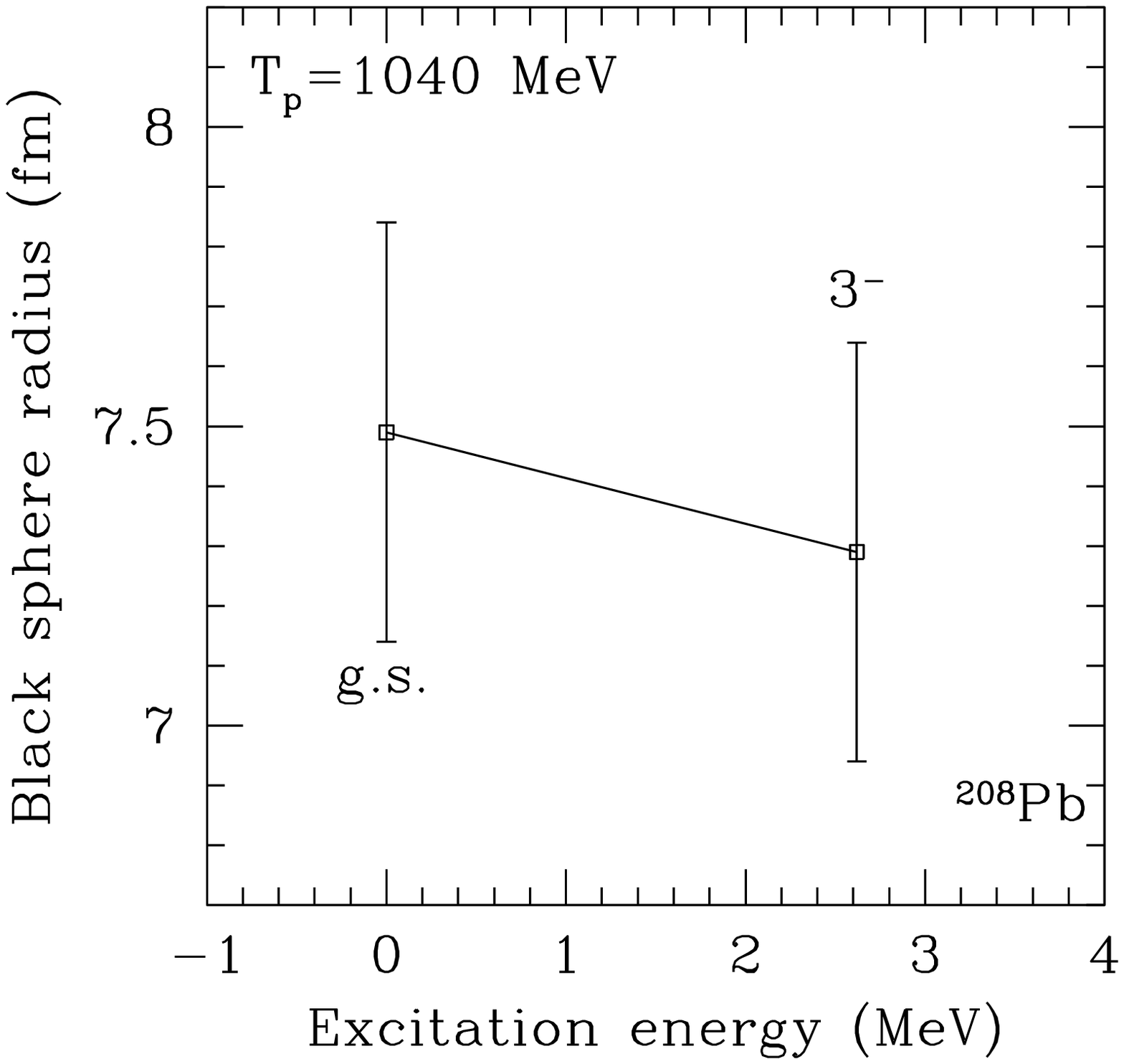}
 \end{minipage}
\end{center}
\caption{(Color online) The black sphere radius as a function of 
the excitation energy.  The target nucleus is $^{12}$C, $^{58,60,62,64}$Ni,
and $^{208}$Pb.  For Ni isotopes, the squares (with line), circles 
(with short-dashed line), triangles (with long-dashed line), and 
crosses (with dash-dotted line) denote the results for $^{58}$Ni, $^{60}$Ni, 
$^{62}$Ni, and $^{64}$Ni, respectively.  
} 
\end{figure}

     To see the systematic behavior of the black sphere radii, we plot
the radii as function of $E_{\rm ex}$ in Fig.\ 2.  We find that
the black sphere radius tends to increase with $E_{\rm ex}$, with a
few exceptions in which case the black sphere radius decreases with 
$E_{\rm ex}$ in its central value but can be regarded as unchanged allowing 
for the error bars.  This is consistent with the behavior of the 
transition radii obtained systematically from electron inelastic scattering 
off $^{208}$Pb \cite{Friedrich}.

     Recall that the black sphere radius $a_l$ is related to the 
transition density and thus expected to lie between the sizes in the ground 
state and in the excited state.  It is thus reasonable that the differences
between $a$ and $a_l$ for the low-lying excited states of interest here are 
generally small.  The only exception that we discovered
is the $^{12}$C $0^+$ state (the Hoyle 
state), for which $a_0/a\simeq1.16$.  The nucleus in the Hoyle state is 
thus expected to be larger than that in the ground state by more than 16 \%, 
a feature consistent with the $\alpha$-clustering picture of the Hoyle state
\cite{Takashina}.

     In summary, we generalized the black-sphere method for deducing the
size of the ground-state nuclei from proton-nucleus elastic scattering data 
to the case of proton-nucleus inelastic scattering data and provided 
implications for the size of excited nuclei.  In the present analysis, 
we confined ourselves to even-even nuclei.  Extension to odd nuclei
would be straightforward if the conventional collective model works
\cite{Blair}.  It is also important to note that the validity of the 
inelastic Fraunhofer diffraction formula used here tends to decrease with 
increasing scattering angle.  We assume that the peak selected for 
determination of $a_l$, as shown in Fig.\ 1, is in the valid regime, 
which could be checked by microscopically clarifying a relation
between the transition density and $a_l$ \cite{Hashimoto}.  We hope that 
the present analysis could develop into a systematic drawing of the 
black-sphere radii of isomers and nuclei in other characteristic excited 
states over a chart of the nuclides.

\acknowledgments

     We acknowledge the members of Japan Charged-Particle Nuclear Reaction 
Data Group for kindly helping us collect various data sets and 
H. Kondo for useful discussions.

\end{document}